\documentclass[12pt]{iopart}

\usepackage{graphicx}

\begin{document}

\title{Doping Dependence of Gap Inhomogeneities  at Bi$_{2}$Sr$_{2}$CaCu$_{2}$O$_{8+\delta}$ Surfaces}

\author{Wei Chen$^{1}$\footnote{Present address: Max Planck Institute for Solid State Research, Heisenbergstra$\ss$e 1, 70569 Stuttgart, Germany}, Marc Gabay$^{2}$, and P. J. Hirschfeld$^{3}$}
\address{$ˆ1$School of Physics, University of New South Wales,
Sydney 2052, Australia}
\address{$ˆ2$Laboratoire de Physique des Solides, Universit\'{e} Paris-Sud 11, Centre d'Orsay, 91405 Orsay Cedex, France}
\address{$ˆ3$Department of Physics, University of Florida, Gainesville, FL 32611, USA}
\ead{w.chen@fkf.mpg.de}
\begin{abstract}
{ We study  the inhomogeneity of the electronic pairing gap observed by STM near the surface of Bi$_{2}$Sr$_{2}$CaCu$_{2}$O$_{8+\delta}$ to be correlated with interstitial O defects. We treat the problem in a slave boson mean field theory of a disordered  $t-t^{\prime}-J$ model, and identify three aspects of the O defects related to the inhomogeneity: (1)the superexchange interaction is locally enhanced in their vicinity, which enhances the local pairing gap and reduces the coherence peak; (2)they donate holes into CuO$_2$ plane, which reduces the spinon density of states of  and hence the average gap at large doping; (3)holes are locally attracted to the vicinity of oxygen defects, which causes impurity bound state and further reduces the coherence peak. The interplay of these mechanisms explains simultaneously the locally enhanced pairing gap around oxygen defects, and the reduction of average gap as increasing oxygen concentration.}
\end{abstract}
\submitto{\NJP}
\maketitle

\section{Introduction}

The effect of disorder in high-$T_{c}$ cuprates, either intentionally introduced or present as a result of  the doping process, has always been crucial to their practical applications. For several classes of cuprates, Scanning Tunneling Microscopy(STM) has been particularly useful to clarify how disorder affects superconductivity \cite{Kapitulnik1,davisinhom1,davisinhom2,McElroy05,Gomes07,Pasupathy08}. In the frequently studied  Bi$_{2}$Sr$_{2}$CaCu$_{2}$O$_{8+\delta}$(BSCCO) system, where disorder is induced  primarily by interstitial oxygens which dope the system, STM data show a very pronounced inhomogeneity of superconducting gap over the sample surface.   In particular, the gap magnitude is significantly enhanced in the vicinity of oxygen defects \cite{McElroy05}. A phenomenological model was proposed by Nunner et al \cite{Nunner05} based on the assumption that the pairing interaction was locally enhanced in the vicinity of oxygen defects, which successfully captured the feature of locally enhanced superconducting gap, and the reduction of coherence peak in the large gap regions, as well as other statistical correlations.   These authors pointed out that such an enhancement could be understood from a strong coupling perspective if
 the local superexchange $J({\bf r})$ was increased by the presence of a nearby dopant.  Maska et al \cite{Maska07} then showed that within a single-band Hubbard model, local variations in site energies as might be due to a dopant perturbation would always lead to an enhanced $J$ within second-order perturbation theory, but similar calculations within a three-band Hubbard model \cite{Foyevtsova09} indicated that the sign of $J$ modulation was not universal.  Johnston et al \cite{Johnston09} found a suppression of the superexchange in Cu-O cluster calculations with point charge estimates of Madelung energies, but pointed out that enhancement could occur if the electron-phonon coupling to $c$-axis oxygen modes were accounted for.  From a weak-coupling perspective, Foyevtsova et al \cite{Foyevtsova10} considered the change in local electronic structure due to an interstitial O in first principles calculations of BSCCO, and used the results to calculate the spatially averaged spin fluctuation interaction in undoped and doped system; the enhanced pairing in the doped system was similarly attributed to local enhancement around O interstitials.
 A somewhat different approach has been pursued by other groups \cite{Mori08,Khaliullin10}, who have discussed the large polarizability of the
  O interstitial as possibly modulating the overall polarizability of the Cu-O complex and hence the pairing.

%

Although the inhomogeneous superconducting gap and several important correlations among STM observables are nicely reproduced by the locally enhanced pairing interaction picture at optimal doping, a puzzle remained: if each oxygen defect locally enhances superconducting gap, why does the average gap decrease with increasing oxygen concentration? This feature has been quantified in \cite{Gomes07}, where empirically not only is the average gap  found to be decreasing with oxygen concentration, but the spatial distribution of superconducting gap is observed to be narrower at large doping.
This apparent discrepancy indicates that additional ingredients are necessary in order to understand the doping dependence of the inhomogeneous superconducting gap.
 A natural assumption is that at least one of these additional ingredients must be  global electronic correlations, which affects the average superconducting gap over a rather large scale, but still allows the local gap to be enhanced around each oxygen defect.

In this work we employ slave boson mean field theory to study the inhomogeneous pairing gap on the surface of BSCCO, and show that besides the locally enhanced pairing interaction, the holes donated by oxygen defects have a dramatic effect on the pairing gap. We find that at large oxygen concentration, the hole doping level increases accordingly, which reduces the density of states(DOS) of  spinons and causes the average gap to drop. The advantage of slave boson mean field theory is that it treats disorder and strong correlations on equal footing, since the inclusion of strong correlations is necessary to explain the reduction of DOS. A theory of this kind has been applied to study YBa$_{2}$Cu$_{3}$O$_{6+x}$(YBCO) in the presence of a point-like repulsive impurity \cite{Gabay08}, where the repulsion of holes and the reduction of the fermionic bandwidth in the vicinity of the impurity influence charge and spin response significantly, yielding a Nuclear Magnetic Resonance(NMR) spectrum  consistent with experiments. In essence, oxygen defects in BSCCO serve as impurities of the opposite kind: their Coulomb potential  attracts holes which, in the slave boson description, promotes the enhancement of the fermionic bandwidth. We will demonstrate that the enhancement of fermionic bandwidth serves as another mechanism to reduce the coherence peak in the large gap regions.

 A similar method \cite{Wang02} has been proposed previously to study the effect of oxygen defects in BSCCO, where 
 screening of the ionic Coulomb potential is studied in detail, and is argued to be the origin of electronic imhomogeneity. However, without a locally enhanced pairing interaction, the superconducting gap is found to be anticorrelated with oxygen defects, in contrast to the observation of STM. It was also proposed that postulating two types of oxygen dopants, one playing the role of Coulomb defect and the other purely as controlling the doping,  can resolve the O-gap anticorrelation issue \cite{Zhou07}. However, based on various microscopic calculations \cite{Mori08,Khaliullin10,Foyevtsova09,Foyevtsova10,Maska07,Johnston09} that support the possible enhancement of the local pairing interaction by oxygen defects, we believe that it is important to include this effect in a phenomenological way, and treat each oxygen defect  as both a Coulomb center and a source of charge. We will show that the positive correlation between pairing gap and oxygen defects is recovered by properly accounting for various energy and length scales in the problem. It is also proposed that slave boson treatment naturally enhances the pairing interaction around oxygen defects, without any additional microscopic mechanisms such as spatially varying atomic level \cite{Zhu05}. The argument is based on the renormalization of hopping integral $t_{ij}$ and its relation with the superexchange interaction $J_{ij}=4t_{ij}^{2}/U$. However, this mechanism gives a very small enhancement of $J_{ij}$, since we found that the hopping integral is only renormalized by about $20\%$ near oxygen defects, which cannot account for the large gap inhomogeneity observed in STM. The inclusion of a phenomenological, locally enhanced pairing interaction in the framework is therefore still necessary  in the development of the theory.

The structure of the paper is as follows: in Sec. II we revisit the enhanced pairing interaction picture in \cite{Nunner05}, and study the change of local bandwidth due to strong correlations phenomenologically. This section essentially illustrates all the physical effects that occur in the vicinity of each oxygen defect. In Sec. III we introduce the slave boson mean field treatment for homogeneous cuprates, and demonstrate the well-known reduction of average gap due to hole doping. In Sec. IV, oxygen defects are introduced into the slave boson treatment to study disorder and correlations on an equal footing. Sec. V concludes with our results.

\section{phenomenological model of a single oxygen defect}

In this section, we follow the strategy of \cite{Nunner05} to study a single oxygen defect within a Hartree-Fock-Gor'kov mean field approach. The motivation is to clarify the influence of oxygen defects on the local density of states(LDOS), in the simplest non-self-consistent manner. In comparison with the fully self-consistent treatment in Sec. IV, only the fermionic degrees of freedom are considered in this section, since they are essential for ${\it local}$ pairing, and the effect of strong correlations is 
discussed phenomenologically. We consider the following impurity Hamiltonian
\begin{eqnarray}
H&=&\sum_{ij\sigma}\left(-t_{ij}-dt_{ij}\right)c_{i\sigma}^{\dag}c_{j\sigma}+\sum_{i\sigma}\left(V_{i}-\mu\right)n_{i\sigma}
\nonumber \\
&+&\sum_{\langle ij\rangle}\left(\Delta_{ij}+d\Delta_{ij}\right)c_{i\uparrow}^{\dag}c_{j\downarrow}^{\dag}+h.c.
\label{Nunner_Hamiltonian}
\end{eqnarray}
For simplicity and consistency with \cite{Nunner05,Fangetal2006}, in this section we assume that the oxygen defect is located on top of a random Cu site (this is not quite correct \cite{He06}), and we choose $t=1=400$meV for nearest-neighbor and $t^{\prime}=-0.3$ for the next-nearest-neighbor hopping in this section. The homogeneous pairing amplitude is $\Delta_{ij}=\Delta_{0}\left(\delta_{i=r}\delta_{j=r\pm x}-\delta_{i=r}\delta_{j=r\pm y}\right)$. In \cite{Nunner05}, it is also shown that if only the electronic(fermionic) degrees of freedom are accounted for, the usual association of an on-site Coulomb potential $V_{i}$ to each oxygen defect gives a form of the LDOS, including a highly inhomogeneous low-energy LDOS, that is inconsistent with STM experiments. Here we focus on the effect of hopping impurities $dt_{ij}$ and the off-diagonal(pairing channel) impurities $d\Delta_{ij}$, and assume that they only differ from the bulk value on the four bonds connected to impurity site $r$, $dt_{ij}=dt_{ji}=dt\left(\delta_{i=r}\delta_{j=r\pm x,r\pm y}\right)$, and $d\Delta_{ij}=d\Delta_{ji}=d\Delta\left(\delta_{i=r}\delta_{j=r\pm x}-\delta_{i=r}\delta_{j=r\pm y}\right)$.

\begin{figure}[h]
\centering
\includegraphics[width=0.6\columnwidth,clip=true]{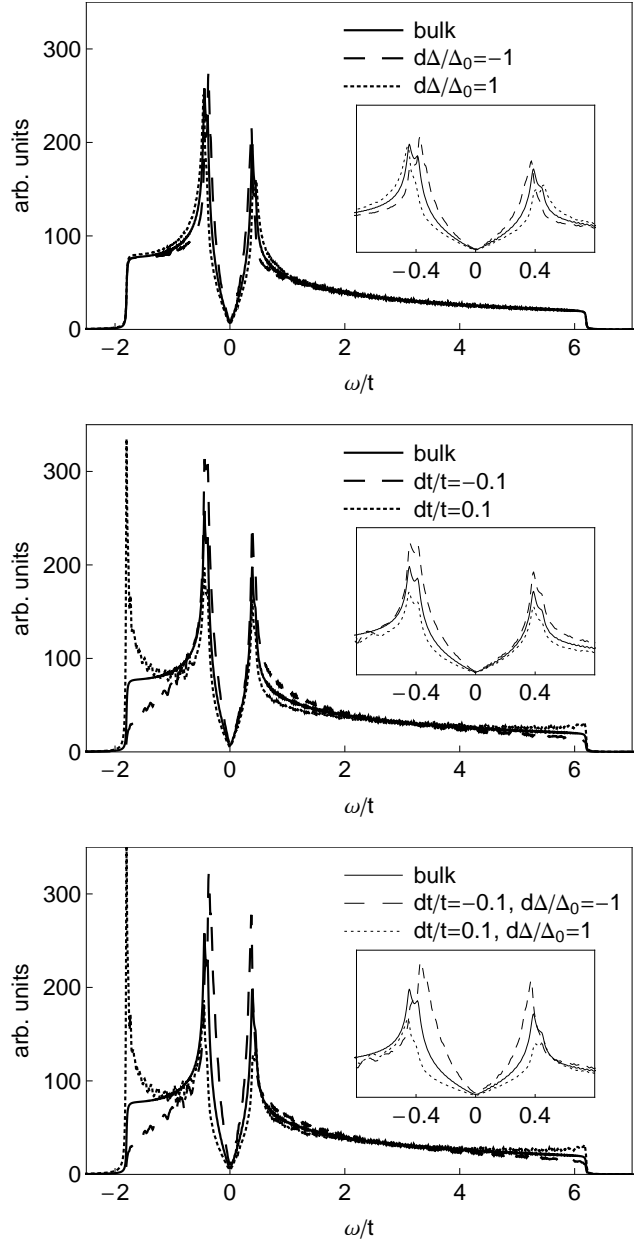}
\caption{LDOS on the impurity site of the phenomenological model described by equation (\ref{Nunner_Hamiltonian}), with different impurity models: (a)Pairing enhancement/reduction $d\Delta/\Delta_{0}=\pm 1$ on four bonds connected to the impurity site, (b)hopping enhancement/reduction $dt/t=\pm 0.1$ on four bonds, and (c)both pairing and hopping enhancement/reduction.  }
\label{fig:LDOS_nonselfconsistent}
\end{figure}

For comparison with \cite{Nunner05,Fangetal2006}, we first perform simulations for a single off-diagonal impurity. As shown in figure \ref{fig:LDOS_nonselfconsistent}(a), one clearly sees a enlarged gap and reduced coherence peak for the case of local pairing enhancement $d\Delta/\Delta_{0}=1$. The off-diagonal impurity case obviously captures the enhancement of the gap by oxygen defects.
 We then perform simulations for the hopping impurity, for both hopping enhancement and reduction cases. The motivation comes from the slave boson treatment of BSCCO surface, where the spinon hopping amplitude is enhanced near oxygen defects due to accumulation of holons, as will be addressed in Sec. IV. Here we treat this effect at the phenomenological level by using a hopping amplitude locally modulated by roughly $dt/t\sim 10\%$ as found by the microscopic theory discussed in Sec. IV. Figure \ref{fig:LDOS_nonselfconsistent}(b) shows the simulation for both hopping enhancement and reduction on four bonds connected to the oxygen defect. The LDOS on the impurity site clearly indicates that hopping enhancement reduces the coherence peak and vice versa, while the gap size remains unchanged. The LDOS calculated over a wider energy range, as shown in figure \ref{fig:LDOS_nonselfconsistent}(b), demonstrates that the reduced coherence peak is due to additional bound states formed on  the top and the bottom of the band. Since the overall spectral weight has to be conserved, the formation of bound state reduces the spectral weight away from the band edge, particularly at the coherence peak.

The simulation that contains both pairing and hopping enhancement is shown in figure \ref{fig:LDOS_nonselfconsistent}(c), where one sees that the superconducting gap is enlarged as expected, and the coherence peak is further reduced compared to the case that contains only an off-diagonal impurity. On the other hand, if the impurity locally reduces both the pairing and hopping interactions, then the gap is reduced and the coherence peak height increases. This simulation 
suggests that the change in the local hopping amplitude, whose origin can be traced back to the holon accumulation due to strong correlations, plays an important role in the shape of STM spectrum, particularly the weight of coherence peak. The calculation that associates change of hopping amplitude to correlation effects will be demonstrated in Sec. IV.

\section{slave boson mean field theory in a homogeneous cuprate}

Before we introduce the disorder model for BSCCO surface in the next section, we present here the slave boson mean field model of a homogeneous cuprate. One can regard the phenomenological impurity model in Sec. II as a demonstration of how oxygen defects affect STM spectrum {\it locally}, while the mean field theory in this section demonstrates how oxygen defects change the average gap value {\it globally}. 
In the slave boson mean field treatment at finite doping, spin and charge degrees of freedom can be separated but mutually renormalize each other. The pairing gap is described by the pairing of spinons, which relies on the LDOS of spinon band that in turn is sensitive to hole doping.

We begin with the $t-t^{\prime}-J$ model treated with the slave boson method
\begin{eqnarray}
{\cal H}&=&\sum_{ij\sigma}-t_{ij}\;b_{i}b_{j}^{\dag}f_{i\sigma}^{\dag}f_{j\sigma}+\sum_{\langle ij\rangle}J\left({\vec S}_{i}\cdot{\vec S}_{j}-\frac{1}{4}n_{i}^{f}n_{j}^{f}\right)
\nonumber \\
&-&\mu_{f}\sum_{i\sigma}n_{i\sigma}^{f}-\mu_{b}\sum_{i}n_{i}^{b}
+\sum_{i\sigma}\lambda_{i}\left(n_{i\sigma}^{f}+n_{i}^{b}-1\right)\;.
\nonumber \\
&&
\label{SB_Hamiltonian}
\end{eqnarray}
We choose $t=400$meV$=1$ and next-nearest-neighbor hopping $t^{\prime}=-0.4$ , a slightly different value from Sec. II such that the van Hove singularity always locates close to but lower than the pairing gap. The exchange coupling $J=72$meV$=0.18$ is smaller than the well accepted value $J\approx 130$meV, hence one expects a smaller gap in the homogeneous limit. However, we found that the local enhancement of $J$ due to oxygen defects restores the pairing strength, and results in a pairing gap comparable to the STM data, as will be demonstrated in Sec. IV.

In equation (\ref{SB_Hamiltonian}), the electron operator has been decomposed into the spin(spinon) and charge(holon) component, $c_{\i\sigma}=f_{i\sigma}b_{i}^{\dag}$. $n_{i\sigma}^{f}=f_{i\sigma}^{\dag}f_{i\sigma}$ and $n_{i}^{b}=b_{i}^{\dag}b_{i}$ represent the spinon and holon density, respectively. In the homogeneous case, the nondouble occupancy constraint $\sum_{\sigma}f_{i\sigma}^{\dag}f_{i\sigma}+b_{i}^{\dag}b_{i}=1$ can be satisfied by adjusting spinon and holon chemical potential independently, so the Lagrangian multiplier $\lambda_{i}$ is practically
irrelevant.
We apply the following mean field decoupling to the Heisenberg term
\begin{eqnarray}
&&\vec{S}_{i}\cdot\vec{S}_{j}=S^{z}_{i}S^{z}_{j}-\frac{1}{2}D^{\dag}_{ij}D_{ij}+\frac{1}{2}\sum_{\sigma}n_{i\sigma}^{f}n_{j-\sigma}^{f}\;.
\label{SSdecoupling}
\end{eqnarray}
where
\begin{eqnarray}
&&{\vec S}_{i}=\frac{1}{2}f_{i\alpha}^{\dag}{\vec \sigma}_{\alpha\beta}f_{i\beta}\;,
\nonumber \\
&&D_{ij}^{\dag}=\sum_{\sigma}\sigma f_{i-\sigma}^{\dag}f_{j\sigma}^{\dag}\;,
\nonumber \\
\label{S_D_order_parameter}
\end{eqnarray}
The Hamiltonian is then
\begin{eqnarray}
{\cal H}&=&\sum_{ij\sigma}-t_{ij}\;b_{i}b_{j}^{\dag}f_{i\sigma}^{\dag}f_{j\sigma}+\sum_{\langle ij\rangle}J\left(S^{z}_{i}S^{z}_{j}-\frac{1}{2}D^{\dag}_{ij}D_{ij}\right)
\nonumber \\
&-&\mu_{f}\sum_{i\sigma}n_{i\sigma}^{f}-\mu_{b}\sum_{i}n_{i}^{b}
+\sum_{i\sigma}\lambda_{i}\left(n_{i\sigma}^{f}+n_{i}^{b}-1\right)\;.
\nonumber \\
&&
\end{eqnarray}
Notice that the spinon density-density term $n_{i}^{f}n_{j}^{f}/4$ is cancelled. The resulting Hamiltonian is slightly different from other slave boson 
decoupling schemes \cite{Ubbens92,Lee92}. However, we do emphasize the importance of keeping the $S_{i}^{z}S_{j}^{z}$ term. A mean field decoupling of this kind shows that $S_{i}^{z}S_{j}^{z}$ accounts for the magnetic correlations that cause impurity induced magnetization in the normal state \cite{Gabay08}. Evidence from neutron scattering indicates an induced magnetization in the center of a vortex \cite{Lake02}, which suggests that magnetic correlations also exist in the superconducting state. We therefore choose to adopt the mean field decoupling in equation (\ref{SSdecoupling}), as it is crucial to other calculations involving magnetic field.

Renormalization of the holon bandwidth is associated with the following operator
\begin{eqnarray}
&&K_{ij\sigma}=f_{i\sigma}^{\dag}f_{j\sigma}\;,
\label{K_C_order_parameter}
\end{eqnarray}
while the renormalization of the spinon bandwidth is simply associated with $b_{i}$ at low temperature. In the mean field treatment, operators in equations (\ref{S_D_order_parameter}) and (\ref{K_C_order_parameter}) are replaced by their ensemble average
\begin{eqnarray}
&&\langle S_{i}^{z}\rangle=m_{i}\;,\nonumber \\
&&\langle D_{ij}^{\dag}\rangle=d_{ij}^{\ast}\;,\nonumber \\
&&\langle K_{ij\sigma}\rangle=\chi_{ij\sigma}\;,\nonumber \\
&&\langle b_{i}^{\dag}\rangle=b_{i}^{\ast}\;,\nonumber \\
&&\langle b_{i}^{\dag} b_{j}\rangle=Q_{ij}\;.\label{MF_order_parameters}
\end{eqnarray}
The nonzero value of $\langle b_{i}^{\dag}\rangle$ in equation (\ref{MF_order_parameters}) implies Bose-Einstein condensation(BEC) in the holon sector. At temperature relevant to the STM measurement that we primarily compare with \cite{Gomes07}, roughly $T_{STM} < 50$K, we find that the holon ground state takes more than $99\%$ of the hole population at any doping larger than $5\%$. Thus one can safely assume that bosonic order parameters are entirely determined by ground state wave functions, and we expect $\langle b_{i}^{\dag} b_{j}\rangle=\langle b_{i}^{\dag}\rangle \langle b_{j}\rangle=Q_{ij}$ (see below). The usual mean field procedure is applied to the above formalism \cite{Lee92}, which yields the following Lagrangian for spinons and for bosonic excitations
\begin{eqnarray}
{\cal L}&=&\sum_{i\sigma}f_{i\sigma}^{\ast}\left(\partial_{\tau}-\mu_{f}\right)f_{i\sigma}
+\sum_{i}b_{i}^{\ast}\left(\partial_{\tau}-\mu_{b}\right)b_{i}
\nonumber \\
&+&\sum_{i}\lambda_{i}\left(\sum_{\sigma}f_{i\sigma}^{\ast}f_{i\sigma}+b_{i}^{\ast}b_{i}-1\right)
+\sum_{\langle ij\rangle}JS_{i}^{z}m_{j}
\nonumber \\
&+&\sum_{ij\sigma}-t_{ij}Q_{ij}f_{i\sigma}^{\ast}f_{j\sigma}
+\sum_{ij}\left(-t_{ij}\sum_{\sigma}\chi_{ij\sigma}\right)b_{i}^{\ast}b_{j}
\nonumber \\
&+&\sum_{\langle ij\rangle}-\frac{J}{2}d_{ij}^{\ast}D_{ij}+h.c.
\end{eqnarray}
In the homogeneous $d-$wave superconducting state, the following ansatz is made for the order parameters
\begin{eqnarray}
&&\chi_{\langle ij\rangle\sigma}=\chi_{\langle ij\rangle\sigma}^{\ast}=\chi^{0}\;,
\nonumber \\
&&\chi_{\langle\langle ij\rangle\rangle\sigma}=\chi_{\langle\langle ij\rangle\rangle\sigma}^{\ast}=\chi^{0\prime}\;,
\nonumber \\
&&Q_{\langle ij\rangle\sigma}=Q_{\langle ij\rangle\sigma}^{\ast}=Q^{0}\;,
\nonumber \\
&&Q_{\langle\langle ij\rangle\rangle\sigma}=Q_{\langle\langle ij\rangle\rangle\sigma}^{\ast}=Q^{0\prime}\;,
\nonumber \\
&&d_{ij}=d_{ij}^{\ast}=\left\{\begin{array}{ll}d_{0} & {\rm for}\; j=i\pm{\hat x} \\
-d_{0} & {\rm for}\;{j=i\pm{\hat y}}
\end{array}\right.
\nonumber \\
&&m_{i}=0\;,
\end{eqnarray}
where $\langle ij\rangle$ and $\langle\langle ij\rangle\rangle$ denote nearest- and next-nearest-neighbor bonds,
 respectively. The Lagrangian of holon and spinon are then separated. The spinon part has dispersion and  $d-$wave gap given by
\begin{eqnarray}
&&\xi_{\bf k}=-2tQ^{0}\left(\cos k_{x}+\cos k_{y}\right)-4t^{\prime}Q^{0\prime}\cos k_{x}\cos k_{y}-\mu_{f}\;,
\nonumber \\
&&\Delta_{\bf k}=Jd_{0}\left(\cos k_{x}-\cos k_{y}\right)\;,
\nonumber \\
&&H_{f}=\sum_{\bf k}\Phi_{\bf k}^{\ast}\left(\begin{array}{ll}\xi_{\bf k} & \Delta_{k} \\
\Delta_{k} & -\xi_{\bf k}
\end{array}\right)\Phi_{\bf k}\;,
\nonumber \\
&&\Phi_{\bf k}=\left(\begin{array}{l}f_{{\bf k}\uparrow} \\ f_{{\bf -k}\downarrow}^{\ast} \end{array}\right)\;.
\end{eqnarray}
For the holon part, The Hamiltonian is
\begin{eqnarray}
H_{b}&=&\sum_{\bf k}\omega_{\bf k}b_{\bf k}^{\ast}b_{\bf k}\;,
\nonumber \\
\omega_{\bf k}&=&-4t\chi^{0}\left(\cos k_{x}+\cos k_{y}\right)-8t^{\prime}\chi^{0\prime}\cos k_{x}\cos k_{y}-\mu_{b}\;.
\nonumber \\
&&
\label{holon_Hamiltonian}
\end{eqnarray}
The order parameters are solved self-consistently by means of Bogoliubov-de Gennes(BDG) equation with the inclusion of BEC of holons. For the spinon sector, we apply the following spin-generalized Bogoliubov transformation
\begin{eqnarray}
f_{i\uparrow}&=&\sum_{n}u_{n,i\uparrow}\gamma_{n\uparrow}+v_{n,i\uparrow}^{\ast}\gamma_{n\downarrow}^{\dag}
\nonumber \\
f_{i\downarrow}&=&\sum_{n}u_{n,i\downarrow}\gamma_{n\downarrow}+v_{n,i\downarrow}^{\ast}\gamma_{n\uparrow}^{\dag}\;.
\end{eqnarray}
After diagonalization, the spinon order parameters are calculated by wave functions $u_{n,i\sigma}$ and $v_{n,i\sigma}$, and Fermi distribution $f(E_{n\sigma})$
\begin{eqnarray}
\langle f_{i\sigma}^{\dag}f_{j\sigma}\rangle
&=&\sum_{n>0}\left\{u_{n,i\sigma}^{\ast}u_{n,j\sigma}f(E_{n\sigma})\right.
\nonumber \\
&+&\left.v_{n,i\sigma}^{\ast}v_{n,j\sigma}(1-f(E_{n\overline{\sigma}}))\right\}\;,
\nonumber \\
\langle f_{i\uparrow}f_{i+\delta\downarrow}\rangle&=&\sum_{n>0}\left\{u_{n,i\uparrow}v_{n,i+\delta\downarrow}^{\ast}(1-f(E_{n\uparrow}))\right.
\nonumber \\
&&\left.+v_{n,i\uparrow}^{\ast}u_{n,i+\delta\downarrow}(1-f(-E_{n\downarrow}))\right\}\;.
\label{spinonselfconsistent}
\end{eqnarray}
Bosonic order parameters under BEC are calculated by assuming all holons are condensed  in the ground state
\begin{eqnarray}
&&\langle b_{i}\rangle=\sqrt{N_{0}}\alpha_{0,i}\;,
\label{BEC_order_parameters}
\end{eqnarray}
where $\alpha_{0,i}$ is the ground state wave function of equation (\ref{holon_Hamiltonian}), and $N_{0}=xL^{2}$ is total number of holes on an $L\times L$ lattice. The $d-$wave superconducting state is characterized by the region in the phase diagram where $d_{0}\neq 0$ and 
$T<T_{BEC}$. In reality, BEC in slave boson formalism is a rather complicated issue. As pointed out in \cite{Lee92}, if one naively adopts the BEC formalism for 2D interacting bose systems, the condensation temperature is as high as $T_{BEC}\sim O(1000)$K near optimal doping, which is certainly unrealistic. Such a high condensation temperature is properly reproduced if one generalize the BEC treatment, equations (\ref{MF_order_parameters}) and (\ref{BEC_order_parameters}) to finite temperature. To reduce the condensation temperature to a realistic scale, one needs to include the gauge degrees of freedom and their scattering with holons \cite{Lee92}. Nevertheless, we are primarily interested in low temperature STM results deep inside the condensate, $T\ll T_{BEC}$. Therefore the precise value of $T_{BEC}$ does not affect the results in the present work, and our assumption that holons occupy only the ground state, equation (\ref{BEC_order_parameters}), is adequate. For the calculation of gap inhomogeneity, we choose $T=50$K such that it sits reasonably below $T_{c}$ in most of the doping regime, while safely above the temperature limit due to finite cluster size.  STM data at this temperature is also available to compare with. 

\begin{figure}[h]
\centering
\includegraphics[width=0.8\columnwidth,clip=true]{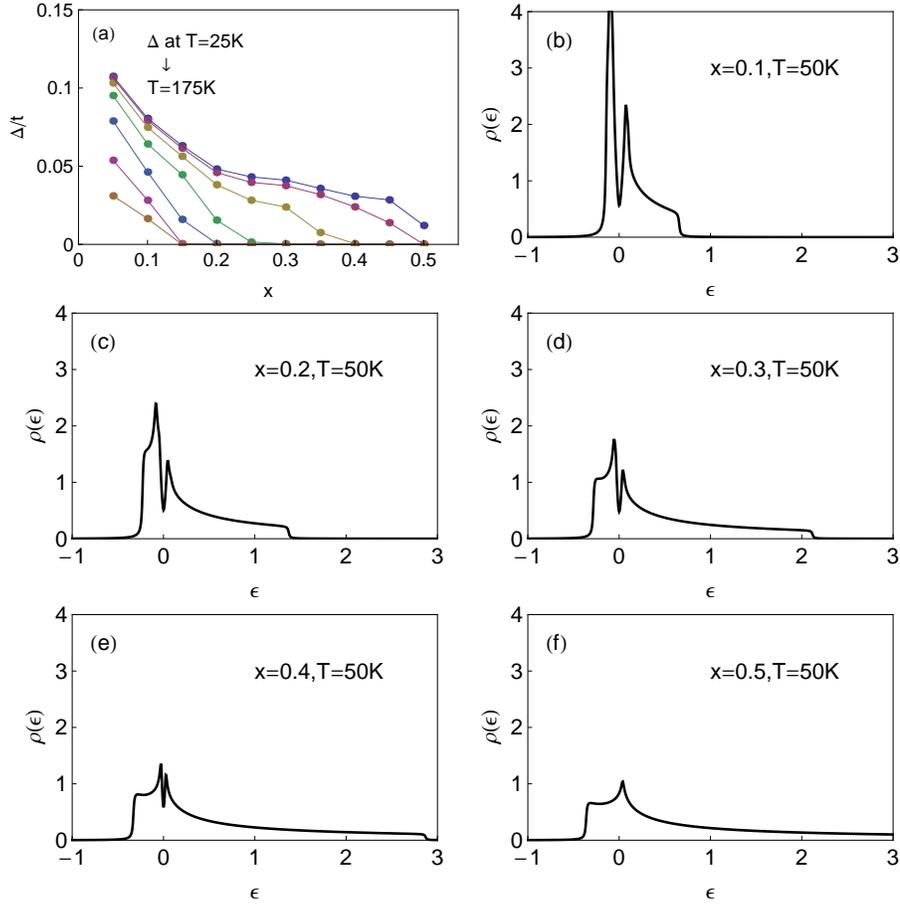}
\caption{(a) Pairing gap $\Delta$ versus temperature and doping $x$. (b)$\sim$(f) Homogeneous DOS at different doping level, at temperature $T=50$K. }
\label{fig:gap_LDOS_vs_x}
\end{figure}

Figure \ref{fig:gap_LDOS_vs_x} shows the pairing gap and DOS as a function of hole doping.  Notice that since holons remain condensed, STM effectively measures the spinon spectrum. The implication of figure \ref{fig:gap_LDOS_vs_x} is straightforward: as doping increases, the spinon bandwidth enlarges, causing reduction of DOS at Fermi surface, hence a smaller pairing gap. The enlargement of the spinon band is due to a larger $Q_{0}$. This is because at low temperatures, $Q_{0}\approx n^{b}=x$, so the spinon bandwidth is roughly proportional to doping.
 Our simulation simply indicates that, although the pairing interaction is enhanced locally around each oxygen defect, the fact that oxygen defect donates holes into CuO$_{2}$ plane can change the spinon DOS dramatically. This serves as another mechanism to determine the pairing gap in a larger length scale, as long as holes are practically not localized. The precise simulation that contains both local and global effects is presented in the next section.

Notice that in order to compare our simulation with real cuprates, the hole doping level has to be rescaled. It is known that slave boson mean field theory of this kind have problems getting both temperature and doping scale correctly \cite{Ruckenstein87,Kotliar88}. We choose our parameters such that the simulation yields a pairing gap close to the value indicated by STM($\Delta\sim 20$meV at optimal doping). The doping level is then rescaled empirically by identifing the doping where superconductivity vanishes $x\approx 0.6$, with that of a real cuprate $p\approx 0.3$, therefore
\begin{eqnarray}
x\approx 2p.
\label{dopingrescale}
\end{eqnarray}
Throughout the article we denote numerical doping scale as $x$, and the true doping scale in cuprates as $p$. One should rather regard equation (\ref{dopingrescale}) as an empirical formula, since the precise correspondence between $x$ and $p$ depends on the mean field decoupling equation (\ref{SSdecoupling}). Nevertheless, the pairing gap and critical temperature at optimal doping($T_{c}\approx 100$K at $p=0.15=x/2$) is very close to the true value in BSCCO.

\section{slave boson mean field treatment of BSCCO surface}
\label{sec:sb_surface}
In this section, we incorporate oxygen defects into slave boson mean field theory studied in Sec. III to simulate the surface of BSCCO. We adopt the phenomenological treatment of \cite{Nunner05} (see also \cite{Zhu05}) that assumes oxygen defects locally enhance pairing strength and attract holes to their vicinity. In contrast to the weak coupling theory in Sec. II, enhancement of the pairing interaction is described by locally varying $J_{ij}$ in the $t-t^{\prime}-J$ model,
\begin{eqnarray}
J_{ij}=J+\delta J\left(h_{i}+h_{j}\right)/2\;,
\label{J_model}
\end{eqnarray}
and the Coulomb interaction yields an additional term in the Hamiltonian described by locally varying potential $V_{i}$
\begin{eqnarray}
&&V_{i}=V_{0}h_{i}\;,
\nonumber \\
&&h_{i}=\sum_{s}\frac{\exp\left(-r_{is}/\lambda\right)}{r_{is}}\;,
\label{V_model}
\end{eqnarray}
where $s$ is the projected position of the
 oxygen defects into the plane. The precise value of $\delta J$, $V_{0}$, and $\lambda$ will be addressed later in the discussion of different energy and length scales. With all these effects together, the Hamiltonian in the slave boson language reads
\begin{eqnarray}
H_{f}&=&\sum_{ij\sigma}-t_{ij}Q_{ij}f_{i\sigma}^{\dag}f_{j\sigma}
+\sum_{ij\sigma}-t_{ij}Q_{ij}^{\ast}f_{j\sigma}^{\dag}f_{i\sigma}
\nonumber \\
&+&\sum_{\langle ij\rangle}J_{ij}\;S_{i}^{z}m_{j}
+\sum_{\langle ij\rangle}-\frac{J_{ij}}{2}d_{ij}^{\ast}D_{ij}
+\sum_{\langle ij\rangle}-\frac{J_{ij}}{2}d_{ij}D_{ij}^{\dag}
\nonumber \\
&+&\sum_{i\sigma}\left(-\mu_{f}+\lambda_{i}\right)n_{i\sigma}^{f}\;,
\nonumber \\
H_{b}&=&\sum_{ij\sigma}-t_{ij}\chi_{ij\sigma}^{\ast}b_{i}^{\dag}b_{j}
+\sum_{ij\sigma}-t_{ij}\chi_{ij\sigma}b_{j}^{\dag}b_{i}
\nonumber \\
&+&\sum_{i}\left(V_{i}-\mu_{b}+\lambda_{i}\right)n_{i}^{b}\;.
\label{H_sb_disorder}
\end{eqnarray}
Notice that $V_{i}$ is present only in the bosonic sector, as it only affects charge degrees of freedom in the problem. In the disorder case, $\lambda_{i}$ serves as a local correction to the chemical potential in order to satisfy the nondouble occupancy constraint. To compare with the density inhomogeneity in \cite{Chen10}, we assume that oxygens are located on top of four adjacent Cu sites, although this is slightly different from the most energetically favorable position \cite{He06}. We assume the positions of oxygen defects are truly random, unlike the nonadjacent condition employed in \cite{Chen10}. This is because if the oxygens are assumed to be nonadjacent to each other, then in a 2D square lattice their concentration can at most be $25\%$, whereas a higher oxygen concentration is necessary to achieve higher doping level in the present work.

\begin{figure}[h]
\leavevmode
\includegraphics[width=0.5\columnwidth,clip=true]{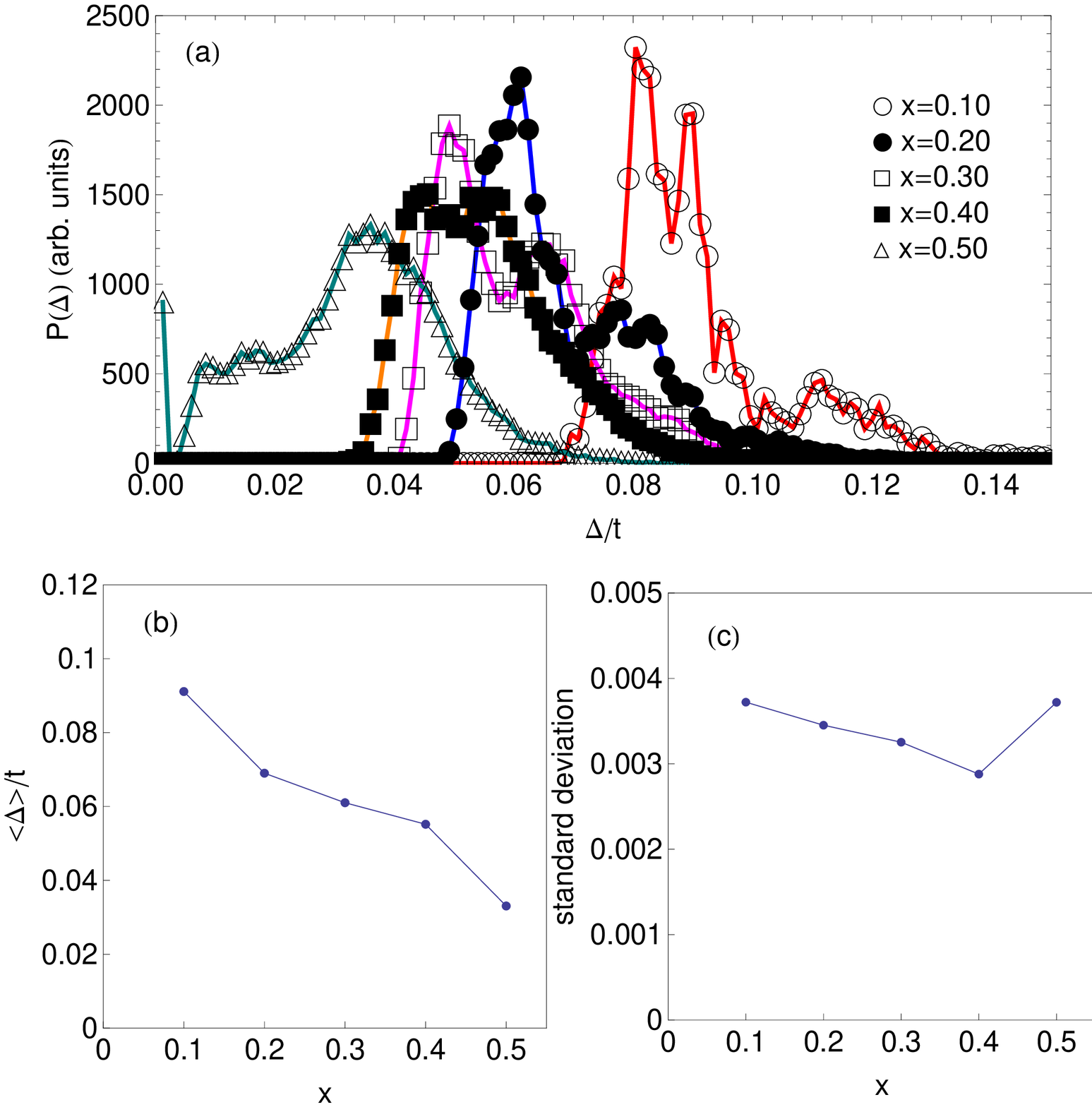}
\includegraphics[width=0.46\columnwidth,clip=true]{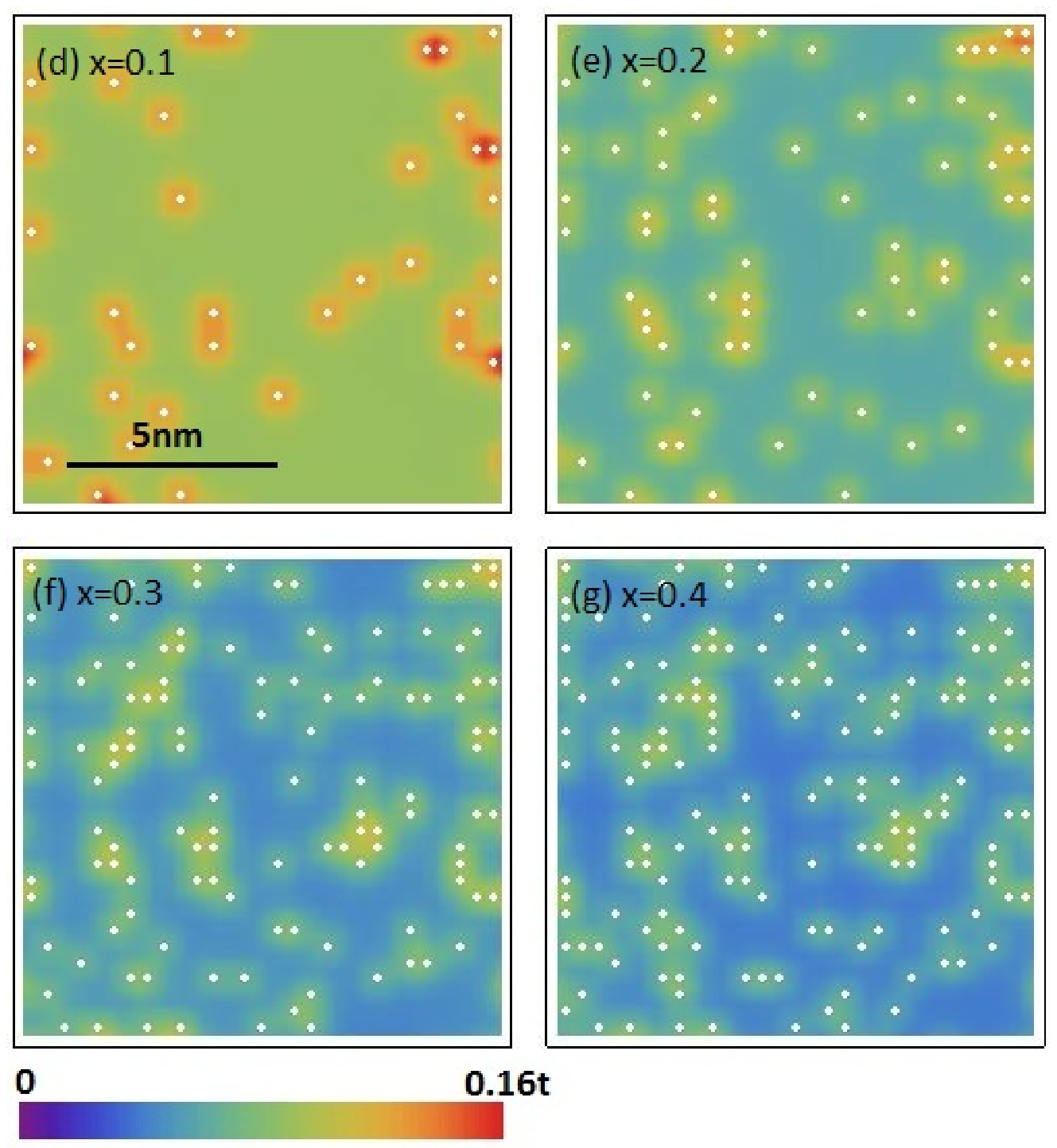}
\caption{(a)Probability distribution $P(\Delta)$ of the pairing gap at different doping, in a $30\times 30$ cluster averaged over 40 impurity configurations. (b)The average gap and (c)the standard deviation calculated from $P(\Delta)$. (d)$\sim$(g) Real space gap map for a specific impurity configuration in different doping levels. Oxygen positions are indicated by white dots. The temperature is fixed at $T=50$K in this figure. }
\label{fig:gap_distribution_map_vs_x}
\end{figure}

Figure \ref{fig:gap_distribution_map_vs_x} shows the probability distribution $P(\Delta)$ and real space map of pairing gap. In figure \ref{fig:gap_distribution_map_vs_x}(a), one sees a fairly broad distribution for the doping regime $0.1\leq x\leq 0.5$, whose
mean shifts to smaller value as doping increases. This is a clear indication that the average pairing gap decreases  with increasing doping, consistent with the observation in STM. Comparing figure \ref{fig:gap_distribution_map_vs_x}(b), where the average pairing gap $\langle\Delta\rangle$ is quantified for each probability distribution, with the homegeneous gap shown in figure \ref{fig:gap_LDOS_vs_x}, it is evident that this $\langle\Delta\rangle$ versus $x$ behavior is inherent from the doping dependence in the homogeneous case. Since the lineshape of $P(\Delta)$ is rather peculiar, we calculate its linewidth by means of standard deviation, as shown in figure \ref{fig:gap_distribution_map_vs_x}(c). In comparison with the STM data \cite{Gomes07}, where the linewidth narrows significantly, about $60\%$ from underdoped to overdoped regime, our simulation shows a narrowing only about $20\%$. We suspect that to correctly capture this $60\%$ reduction of $P(\Delta)$ linewidth, one needs a fine mechanism to control the overlap between superconducting islands as  doping changes, for instance a fine tuning of $\lambda$, or using a more sophisticated impurity model other than equation (\ref{J_model}) and equation (\ref{V_model}). Nevertheless, our calculation demonstrates that within a simplestic impurity model with only one length scale $\lambda$ associated with impurity, we can qualitatively capture the pairing gap inhomogeneity induced by oxygens.  Notice that the linewidth of $P(\Delta)$ increases again in highly overdoped case $x=0.5$. This is because in the absence of oxygen disorder, the homogeneous gap is zero at this doping. Randomly distributed oxygens locally enhance the gap, which causes the distribution to range broadly from zero to $\Delta\approx 0.08$.

 From the real space gap map, figure \ref{fig:gap_distribution_map_vs_x}(d) to (g), the inhomogeneity induced by oxygen defects (white dots) is  clearly visible.  For each doping level,  the pairing gap is locally enhanced near oxygen defects, consistent with the weak coupling approach \cite{Nunner05}. This is particularly evident at low doping level, as shown by red spots in figure \ref{fig:gap_distribution_map_vs_x}(d) for $x=0.1$, where large gap region coincides with the oxygen positions. The oxygen-rich and oxygen-poor regions also give a double line structure to the gap distribution shown in figure \ref{fig:gap_distribution_map_vs_x}(a), particularly for low doping lines from $x=0.1$ to $0.3$. At higher doping, puddles of large gap region start to overlap, which eventually smears out the double line structure at around $x=0.4$.


We now discuss the robustness of the simulation against variation of the parameters, especially interaction strength $\delta J$ and $V_{0}$, and interaction range $\lambda$. We find that in order to have an average pairing gap decreasing with doping, $\lambda$ has to be relatively small, roughly $\lambda\leq 1$. In other words, the interaction has to be modulated
on an atomic scale \cite{Nunner05}.
This can be understood simply because, had the pairing enhancement due to oxygen defects been longer range, it would have overcome the correlation-induced doping effect and enhanced the gap globally.
Secondly, in order to produce a probability distribution that has comparable linewidth to STM data, the pairing enhancement must be roughly the same as the homogeneous superexchange interaction $\delta J\approx J$. This gives an effective coupling $1.34J$ on the nearest Cu-Cu bonds of an oxygen defect. The strength of Coulomb interaction $V_{0}$ is also crucial to pairing gap. We find that if $V_{0}$ is too large, the holon density is highly accumulated near oxygen defects, causing a very low spinon density and a smaller gap, even though pairing interaction is enhanced. As also found by Nunner et al.,
a large value of $V_0$ can destroy the anticorrelation between the coherence peak width and position.   This analysis sets up an upper limit for the Coulomb interaction $V_{0}\leq 0.4$. We choose $V_{0}=0.2$, close to the value at optimal doping obtained by estimation of the dielectric constant \cite{Chen10}, although the value here is taken to be doping independent.

\begin{figure}[htb]
\centering
\includegraphics[width=0.95\columnwidth,clip=true]{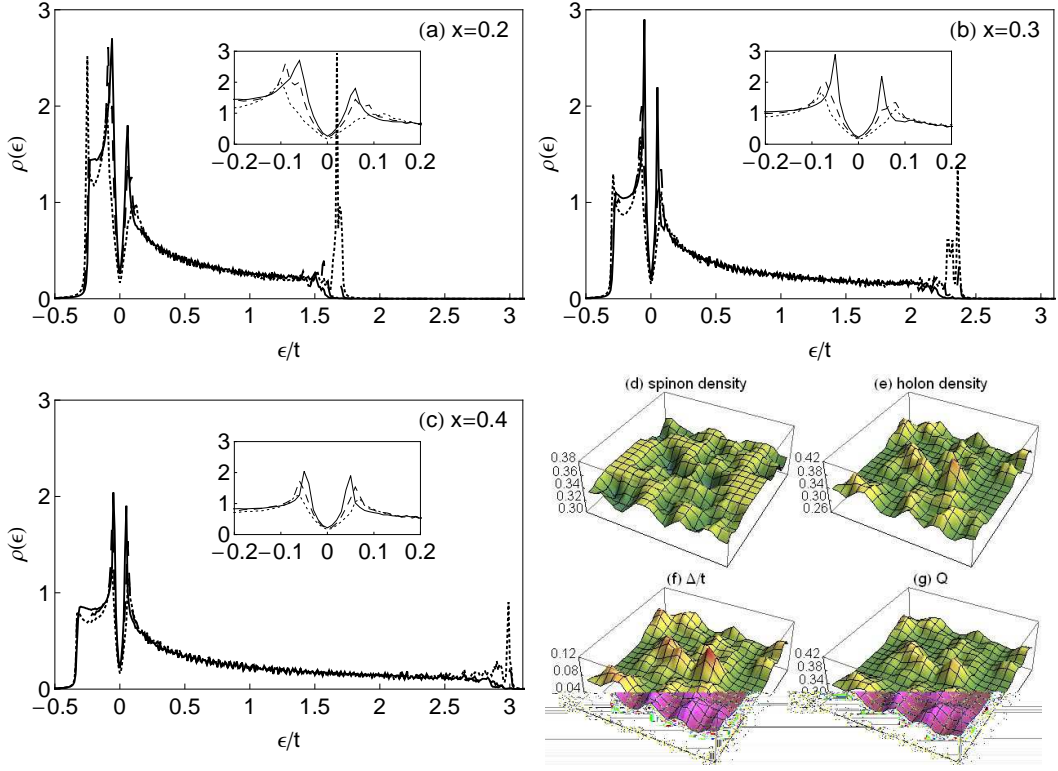}
\caption{(a)$\sim$(c)LDOS at three different average doing. In each figure, solid, dashed, and dotted lines represent typical LDOS at positions with large, medium, and small gap, respectively. Inserts show the LDOS near chemical potential. (d)$\sim$(g)Local order parameters at a specific impurity configuration at $x=0.33$, in a cluster size $7.6$nm$\times 7.6$nm. }
\label{fig:LDOS_order_parameters}
\end{figure}

A simulation of the LDOS and order parameters for a specific disorder configuration is shown in figure \ref{fig:LDOS_order_parameters}. For any doping level examined, the height of the coherence peak is found to decrease with gap size and vice versa, coinciding with \cite{Nunner05} and the STM data. Comparing the $\Delta_{ij}$ map in figure \ref{fig:LDOS_order_parameters}(f) with the $Q_{ij}$ map in figure \ref{fig:LDOS_order_parameters}(g), it is evident that regions with large gap also have large $Q_{ij}$. Similar to the homogeneous case, this is because at relatively low temperatures, bosons are fairly condensed, so $Q_{ij}\approx \sqrt{n_{i}^{b}n_{j}^{b}}$. Thus $Q_{ij}$ is large near oxygen defects because the holon density is rich. Since $Q_{ij}$ enters the spinon Hamiltonian equation (\ref{H_sb_disorder}) as hopping amplitude, enhancement of $Q_{ij}$ means the spinon hopping amplitude is locally enhanced around oxygen defects, as assumed phenomenologically in Section II. On the other hand, the LDOS near the oxygen defects has a very pronounced impurity bound state at band edge, which further reduces coherence peak. From the phenomenological model in Sec. II, it is clear that this bound state is a result of hopping enhancement, i.e. larger $Q_{ij}$ near oxygen defects. Therefore from this calculation of the LDOS, one can see the importance of strong correlations on local bandwidth and how it influences the shape of LDOS.

Figure \ref{fig:LDOS_order_parameters}(e) shows a very significant holon density inhomogeneity. This compares well with \cite{Chen10}, where inhomogeneous charge density on BSCCO surface was studied by means of small hole pockets, For average doping $x=\langle n_{i}^{b}\rangle=0.33$, the holon density varies from $0.3< n_{i}^{b}< 0.4$. However, in the weak coupling mean field approach of \cite{Nunner05}, the hole density inhomogeneity is very moderate, roughly less than $5\%$. This can be understood as the following: in both \cite{Chen10} and  the present work, charge degrees of freedom  are carried  by dilute {\it holes}, which screen Coulomb impurities poorly, while in \cite{Nunner05} it is carried out by {\it electrons}, which screen Coulomb impurities very efficiently. We anticipate that an approach like the present work is particularly accurate for underdoped to optimally doped cuprates, where hole density inhomogeneity is due to poor screening of dilute holes. On the overdoped side, where the system has presumably crossed over to a Fermi liquid, one should perhaps adopt weak coupling approaches like in \cite{Nunner05} to accurately describe the screening by electrons. Notice that the density inhomogeneity in the present work is slightly smaller than in \cite{Chen10}, due to coupling between charge and spin degrees of freedom. This is similar to a previous report where no clear Friedel density oscillation is seen near a point-like repulsive impurity \cite{Gabay08}. This is simply because holons, which respond to the Coulomb interaction of disorder, do not have a characteristic length scale, so its density profile is rather smooth. Due to the nonodouble occupancy constraint, the amplitude of charge inhomogeneity is also suppressed, because holon density is subject to the spinon density that does not respond to Coulomb interaction directly.

\begin{figure}[h]
\centering
\includegraphics[width=0.6\columnwidth,clip=true]{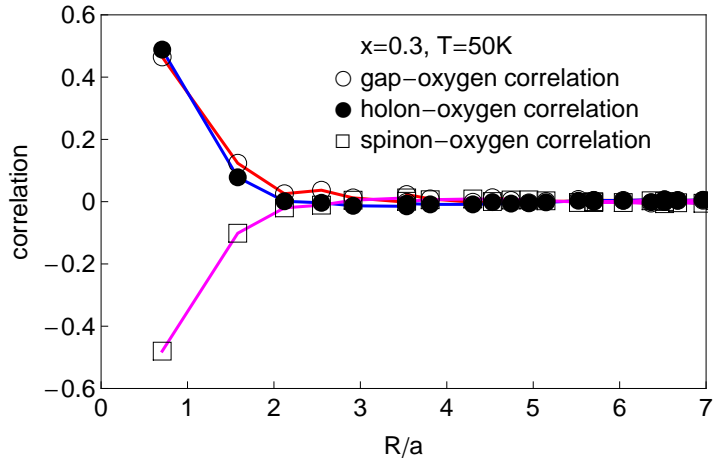}
\caption{Various correlation functions calculated according to \cite{Chen10}, at doping level $x=0.3$ and $T=50$K. }
\label{fig:correlation_function}
\end{figure}

Lastly, we address the issue of various correlation functions related to STM. We define the correlation functions according to \cite{Chen10}, which is consistent with \cite{McElroy05} but defined on a discrete lattice. The results for various correlators are shown in figure \ref{fig:correlation_function}. The oxygen defects are positively correlated with pairing gap. The value of correlation extracts to $\approx 0.7$ for onsite correlation $R=0$, close to the value $\approx 0.4$ extracted from STM \cite{McElroy05}. The holon-oxygen correlation, which is exactly the negative of spinon-oxygen correlation due to the no double occupancy constraint, is always positive, indicating the fact that an oxygen defect attracts holes to its vicinity. The length scale at which correlation drops to zero is roughly $2a$, which is smaller than that reported in \cite{Chen10}. Notice that in \cite{Nunner05}, the potential length scale
$\lambda=0.5$ yields a correlation length in good agreement with
the STM results, whereas the same value of lambda gives a
smaller correlation length in the current approach. We suspect
that this is because spinon correlation functions are subject to
holon correlation functions, and the no double occupancy
constraint reduces the correlation length. Since $\lambda$ is
treated phenomenologically, in principle it can be adjusted
to yield a longer correlation length. However, in the present approach -- where the oxygen concentration is rather high -- our choice of a smaller lambda allows us to keep the pairing enhancement local.

\section{Conclusion}

We have presented a slave boson mean field approach to compare the possible physical effects induced locally by oxygen dopants near the surface of BSCCO, and show how their interplay can explain the pairing gap revealed by STM experiments. Firstly, the enhancement of the pairing gap and reduction of coherence peak near oxygen defects is attributed to locally enhanced pairing interaction via superexchange. Secondly, the average spinon density of states at the chemical potential is reduced at large doping, causing the average gap to decrease, although it remains locally enhanced around oxygen defects. We also found that since holes are attracted to the vicinity of oxygen defects, an impurity bound state is formed at the edge of spinon band, which further causes the reduction of coherence peak, yielding a LDOS shape consistent with STM.

The present work also gives important evidence regarding various length scales and interaction strengths in the cuprates. Since the pairing gap is locally enhanced around oxygen defects but has an average value that decreases with doping, we conclude that the length scale at which the pairing interaction is enhanced is very small $\lambda \approx 0.5a$, although the enhancement $\delta J$ is about the same magnitude as the bare pairing interaction $J$, such that the it gives a significant gap inhomogeneity. On the other hand, the Coulomb interaction associated with the oxygen defects is relatively small compared to the hopping amplitude, such that the accumulation of holes around oxygen defects  does not lead to a reduction of the pairing gap near oxygen defects, although it is large enough to yield a significant charge inhomogeneity in underdoped regime.

\ack

We thank G. Khaliullin, M. Mori, O. P. Sushkov, C. Hamer, J. Oitmaa, and J. C. Davis for stimulating discussions. P.J.H acknowledges support from NSF-DMR-
1005625. Numerical calculation in the present work is done by the facilities in Australian National Computational Infrastructure under project u66.


\section*{References}

\begin{thebibliography}{99}

\bibitem {Kapitulnik1} Howald C, Fournier P and Kapitulnik A 2001
{\it Phys. Rev. B} {\bf 64} 100504(R)

\bibitem{davisinhom1} Pan S H {\it et al} 2001 {\it Nature}
{\bf 413} 282

\bibitem{davisinhom2} Lang K M {\it et al} 2002 {\it Nature} {\bf 415} 412

\bibitem{McElroy05}
McElroy K, Lee J, Slezak J A, Lee D -H, Eisaki H, Uchida S and Davis J C 2005 {\it Science} {\bf 309} 1048 

\bibitem{Gomes07}
Gomes K K, Pasupathy A N, Pushp A, Ono S, Ando Y and Yazdani A 2007 {\it Nature} {\bf 447} 569 

\bibitem{Pasupathy08}
Pasupathy A N, Pushp A, Gomes K K, Parker C V, Wen J,
Xu Z, Gu G, Ono S, Ando Y and Yazdani A, 2008 {\it Science} {\bf 320} 196 

\bibitem{Nunner05}
Nunner T S, Andersen B M, Melikyan A and Hirschfeld P J 2005
{\it Phys. Rev. Lett.} {\bf 95} 177003 

\bibitem{Maska07}
Maska M M, Sledz Z, Czajka K and Mierzejewski M 2007 {\it Phys. Rev. Lett.} {\bf 99} 147006 

\bibitem{Foyevtsova09}
Foyevtsova K, Valentí R and Hirschfeld P J 2009
{\it Phys. Rev. B} {\bf 79} 144424 

\bibitem{Johnston09}
Johnston S, Vernay F and Devereaux T P 2009 {\it Europhys. Lett.} {\bf 86} 37007 

\bibitem{Foyevtsova10}
Foyevtsova K, Kandpal H C, Jeschke H O, Graser S, Cheng H -P, Valentí R and Hirschfeld P J 2010 {\it Phys. Rev. B} {\bf 82} 054514

\bibitem{Mori08}
Mori M, Khaliullin G, Tohyama T and Maekawa S 2008
{\it Phys. Rev. Lett.} {\bf 101} 247003

\bibitem{Khaliullin10}
Khaliullin G, Mori M, Tohyama T and Maekawa S 2010 {\it Phys. Rev. Lett.} {\bf 105} 257005  

\bibitem{Gabay08}
Gabay M, Semel E, Hirschfeld P J and Chen W 2008 {\it Phys. Rev. B} {\bf 77} 165110 

\bibitem{Wang02}
Wang Z, Engelbrecht J R, Wang S, Ding H and Pan S H 2002 {\it Phys. Rev. B} {\bf 65} 064509 

\bibitem{Zhou07}
Zhou S, Ding H and Wang Z 2007 {\it Phys. Rev. Lett.} {\bf 98} 076401 

\bibitem{Zhu05}
Zhu J -X 2005 {\it Preprint} cond-mat/0508646 

\bibitem{Fangetal2006}
Fang A C, Capriotti L, Scalapino D J, Kivelson S A, Kaneko N, Greven M and Kapitulnik A 2006 {\it Phys. Rev. Lett.} {\bf 96} 017007 

\bibitem{He06} 
He Y, Nunner T S, Hirschfeld P J and Cheng H -P 2006 {\it Phys. Rev. Lett.} {\bf 96} 197002 

\bibitem{Ubbens92}
Ubbens M U and Lee P A 1992 {\it Phys. Rev. B} {\bf 46} 8434 

\bibitem{Lee92}
Lee P A and Nagaosa N 1992 {\it Phys. Rev. B} {\bf 46} 5621 

\bibitem{Lake02}
Lake B {\it et al} 2002 {\it Nature} {\bf 415} 299 

\bibitem{Ruckenstein87}
Ruckenstein A E, Hirschfeld P J and Appel J 1987 {\it Phys. Rev. B} {\bf 36} 857 

\bibitem{Kotliar88}
Kotliar G and Liu J 1988 {\it Phys. Rev. B} {\bf 38} 5142 

\bibitem{Chen10}
Chen W, Khaliullin G and Sushkov O P 2011 {\it Phys. Rev. B} {\bf 83} 064514 





\end{thebibliography}
\end{document}